\documentclass[letterpaper, 10pt]{article}

\usepackage{spconf}

\usepackage[T1]{fontenc}
\usepackage[utf8]{inputenc}
\usepackage{times} % assumes new font selection scheme installed
\usepackage[english]{babel}

\usepackage{graphicx}
\usepackage{caption}
\usepackage{subcaption}
\usepackage{tikz}
\usepackage{t1enc, psfrag}
\usepackage[dvips]{epsfig}
\usepackage{epstopdf}

\usepackage{algorithm}
\usepackage[noend]{algpseudocode}
\usepackage{float}

\makeatletter
\def\BState{\State\hskip-\ALG@thistlm}
\makeatother

\makeatletter
\newenvironment{breakablealgorithm}
  {% \begin{breakablealgorithm}
   \begin{center}
     \refstepcounter{algorithm}% New algorithm
     \hrule height.8pt depth0pt \kern2pt% \@fs@pre for \@fs@ruled
     \renewcommand{\caption}[2][\relax]{% Make a new \caption
       {\raggedright\textbf{\ALG@name~\thealgorithm} ##2\par}%
       \ifx\relax##1\relax % #1 is \relax
         \addcontentsline{loa}{algorithm}{\protect\numberline{\thealgorithm}##2}%
       \else % #1 is not \relax
         \addcontentsline{loa}{algorithm}{\protect\numberline{\thealgorithm}##1}%
       \fi
       \kern2pt\hrule\kern2pt
     }
  }{% \end{breakablealgorithm}
     \kern2pt\hrule\relax% \@fs@post for \@fs@ruled
   \end{center}
  }
\makeatother

\newcommand\Algphase[1]{%
\vspace*{-.5\baselineskip}\Statex\hspace*{\dimexpr-\algorithmicindent-2pt\relax}%\rule{\columnwidth}{0.4pt}%
\Statex\hspace*{-\algorithmicindent}\textbf{#1}%
\vspace*{-.7\baselineskip}\Statex\hspace*{\dimexpr-\algorithmicindent-2pt\relax}%\rule{\columnwidth}{0.4pt}%
}

\usepackage{siunitx}

\usepackage{amsmath}
\usepackage{amsfonts}

\usepackage{hyperref}
\usepackage{ragged2e}
\usepackage{ifthen}
\usepackage[left=1.5cm, right=1.5cm, top=1.3cm, bottom=3.5cm]{geometry}
\usepackage{fancyhdr}
\newcommand{\overbar}[1]{\mkern 1.5mu\overline{\mkern-1.5mu#1\mkern-1.5mu}\mkern 1.5mu}
\newcommand{\myforall}[1]{\quad \forall \; {#1}}
\newcommand\emgvec{\mathbf{x}}
\newcommand\emgchan{x}
\newcommand\emgchanvec{\mathbf{x}}
\newcommand\muapmat{\mathbf{H}}
\newcommand\muapvec{\mathbf{h}}
\newcommand\muapscal{h}
\newcommand\sourcevec{\mathbf{s}}
\newcommand\sourcesubvec{\mathbf{s}}
\newcommand\source{s}
\newcommand\R{\mathbb{R}}
\newcommand\nsensor{P}
\newcommand\nsource{Q}
\newcommand{\approxent}[1]{\overbar{#1}}
\newcommand\nsourceapp{\approxent{\nsource}}
\newcommand\sourcesubvecapp{\approxent{\sourcesubvec}}
\newcommand\muapvecapp{\approxent{\muapvec}}
\newcommand\filtlen{M}
\newcommand\ia{p}
\newcommand\ib{q}
\newcommand\tdis{n}
\newcommand\demix{w}
\newcommand\demixvec{\mathbf{w}}

\newcommand\filtlendem{L}
\newcommand\demixedchan{y}

\newcommand\velocity{v}

\title{Convolutive Blind Source Separation on Surface EMG Signals for Respiratory Diagnostics and Medical Ventilation Control}

\name{Herbert Buchner$^{1}$, Eike Petersen$^{2}$, Marcus Eger$^{3}$, and Philipp Rostalski$^{2}$}
\address{\small $^{1}$ Cambridge University, Department of Engineering, Cambridge, UK. {\tt\small hb444@cam.ac.uk}. \\ 
\small $^{2}$ Institute for Electrical Engineering in Medicine, University of L\"ubeck, Germany. \\
{\tt\small \{eike.petersen, philipp.rostalski\}@uni-luebeck.de}. \\
\small $^{3}$ Dr\"agerwerk AG, L\"ubeck, Germany. {\tt\small marcus.eger@draeger.com}.}

\begin{document}

\maketitle

\begin{abstract}
The electromyogram (EMG) is an important tool for assessing the activity of a muscle and thus also a valuable measure for the diagnosis and control of respiratory support. In this article we propose convolutive blind source separation (BSS) as an effective tool to pre-process surface electromyogram (sEMG) data of the human respiratory muscles. Specifically, the problem of discriminating between inspiratory, expiratory and cardiac muscle activity is addressed, which currently poses a major obstacle for the clinical use of sEMG for adaptive ventilation control. It is shown that using the investigated broadband algorithm, a clear separation of these components can be achieved. The algorithm is based on a generic framework for BSS that utilizes multiple statistical signal characteristics. Apart from a four-channel FIR structure, there are no further restrictive assumptions on the demixing system.
\end{abstract}

\section{Introduction}
Blind Source Separation (BSS) plays an important role in processing surface electromyographic (sEMG) measurements~\cite{holobar14} and can be employed to detect and distinguish crosstalk between adjacent muscles~\cite{holobar14}.
Furthermore, several research groups recently have reported the successful removal of electrocardiographic (ECG) interference from EMG signals by means of BSS techniques~\cite{willigenburg12}. 
Both features are crucial in respiratory applications since they allow to discriminate between the activity of inspiratory and expiratory muscles and to remove the strong interference resulting from the electrical activity of the heart.
%Many different techniques have been proposed in the past to tackle the problem of ECG removal, including classical template subtraction and ECG gating~\cite{bartolo96}. Recently, BSS-based approaches for the removal of ECG contamination from the EMG signal have been shown to yield good results in comparison to other techniques~\cite{willigenburg12}.

The mixing model assumed in a particular BSS algorithm is a major determinant of separation success. For the separation of EMG signals, mostly instantaneous mixing models have been employed in the past, despite the fact that these models are only applicable in very specific settings~\cite{leouffre13}. Convolutive mixing models, on the other hand, appear to provide a much more realistic representation of EMG signal propagation paths~\cite{holobar14} but have only rarely been used in applications~\cite{li_embc05, holobar03}, and to the author's knowledge never been applied to the analysis of respiratory EMG.
In contrast to the muscles usually considered for sEMG signal processing, respiratory muscles such as the diaphragm and the abdominal muscles are rather large and flat. This observation distinguishes the proposed respiratory application from various other EMG-based signal processing settings and entails important consequences for the choice of the mixing model, which will be the topic of section~\ref{sec:mixing_models} of this article.

Following the discussion of the different mixing models for BSS of respiratory EMG signals in section~\ref{sec:mixing_models}, we describe a broadband algorithm for convolutive BSS that is based on a generic framework for the simultaneous exploitation of several statistical signal characteristics (TRINICON, TRIple-N Independent component analysis for CONvolutive mixtures)~\cite{book03, book07} in section~\ref{sec:algorithm}. Results of the application of this algorithm to a clinical data set are presented in section~\ref{sec:results}, before section~\ref{sec:conclusion} concludes this article.

\section{Mixing Models for Source Separation of Surface EMG Signals}
\label{sec:mixing_models}
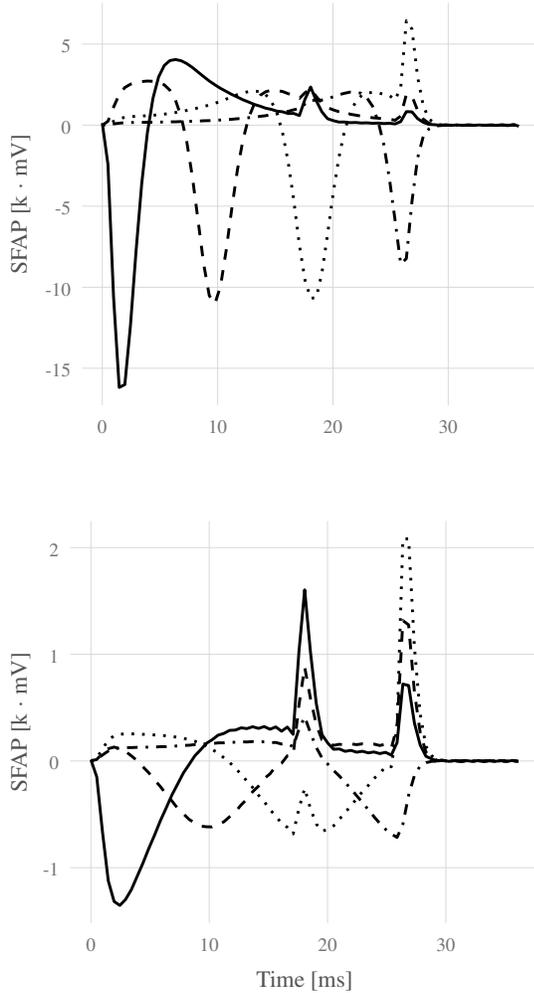
\begin{figure}
  \centering
  \begin{subfigure}[t]{\columnwidth}
    \centering
    {\footnotesize
      % Created by tikzDevice version 0.12 on 2019-03-20 22:22:34
% !TEX encoding = UTF-8 Unicode
\begin{tikzpicture}[x=1pt,y=1pt]
\definecolor{fillColor}{RGB}{255,255,255}
\path[use as bounding box,fill=fillColor,fill opacity=0.00] (0,0) rectangle (216.81,195.13);
\begin{scope}
\path[clip] (  0.00,  0.00) rectangle (216.81,195.13);
\definecolor{drawColor}{RGB}{255,255,255}
\definecolor{fillColor}{RGB}{255,255,255}

\path[draw=drawColor,line width= 0.5pt,line join=round,line cap=round,fill=fillColor] (  0.00,  0.00) rectangle (216.81,195.13);
\end{scope}
\begin{scope}
\path[clip] ( 37.78, 32.50) rectangle (211.12,185.17);
\definecolor{drawColor}{RGB}{255,255,255}
\definecolor{fillColor}{RGB}{255,255,255}

\path[draw=drawColor,line width= 0.5pt,line join=round,line cap=round,fill=fillColor] ( 37.78, 32.50) rectangle (211.12,185.17);
\definecolor{drawColor}{gray}{0.85}

\path[draw=drawColor,line width= 0.3pt,line join=round] ( 37.78, 46.69) --
	(211.12, 46.69);

\path[draw=drawColor,line width= 0.3pt,line join=round] ( 37.78, 77.36) --
	(211.12, 77.36);

\path[draw=drawColor,line width= 0.3pt,line join=round] ( 37.78,108.04) --
	(211.12,108.04);

\path[draw=drawColor,line width= 0.3pt,line join=round] ( 37.78,138.72) --
	(211.12,138.72);

\path[draw=drawColor,line width= 0.3pt,line join=round] ( 37.78,169.39) --
	(211.12,169.39);

\path[draw=drawColor,line width= 0.3pt,line join=round] ( 45.65, 32.50) --
	( 45.65,185.17);

\path[draw=drawColor,line width= 0.3pt,line join=round] ( 89.27, 32.50) --
	( 89.27,185.17);

\path[draw=drawColor,line width= 0.3pt,line join=round] (132.88, 32.50) --
	(132.88,185.17);

\path[draw=drawColor,line width= 0.3pt,line join=round] (176.49, 32.50) --
	(176.49,185.17);
\definecolor{drawColor}{RGB}{0,0,0}

\path[draw=drawColor,line width= 1.1pt,line join=round] ( 45.65,139.47) --
	( 47.78,123.88) --
	( 49.91, 73.53) --
	( 52.04, 39.44) --
	( 54.17, 40.50) --
	( 56.30, 62.92) --
	( 58.43, 91.26) --
	( 60.56,116.59) --
	( 62.69,135.81) --
	( 64.82,148.93) --
	( 66.95,156.97) --
	( 69.08,161.29) --
	( 71.21,163.16) --
	( 73.34,163.56) --
	( 75.47,163.10) --
	( 77.60,162.02) --
	( 79.73,160.55) --
	( 81.86,158.81) --
	( 83.99,157.09) --
	( 86.12,155.42) --
	( 88.25,153.98) --
	( 90.37,152.64) --
	( 92.50,151.49) --
	( 94.63,150.33) --
	( 96.76,149.34) --
	( 98.89,148.30) --
	(101.02,147.46) --
	(103.15,146.56) --
	(105.28,145.91) --
	(107.41,145.14) --
	(109.54,144.69) --
	(111.67,144.02) --
	(113.80,143.78) --
	(115.93,143.13) --
	(118.06,143.15) --
	(120.19,142.34) --
	(122.32,148.55) --
	(124.45,153.16) --
	(126.58,147.96) --
	(128.71,143.75) --
	(130.84,141.16) --
	(132.97,140.45) --
	(135.10,139.89) --
	(137.22,139.91) --
	(139.35,139.66) --
	(141.48,139.73) --
	(143.61,139.53) --
	(145.74,139.60) --
	(147.87,139.43) --
	(150.00,139.50) --
	(152.13,139.33) --
	(154.26,139.44) --
	(156.39,139.19) --
	(158.52,140.05) --
	(160.65,143.83) --
	(162.78,143.75) --
	(164.91,141.20) --
	(167.04,139.80) --
	(169.17,139.00) --
	(171.30,138.88) --
	(173.43,138.69) --
	(175.56,138.78) --
	(177.69,138.66) --
	(179.82,138.78) --
	(181.94,138.65) --
	(184.07,138.78) --
	(186.20,138.65) --
	(188.33,138.79) --
	(190.46,138.63) --
	(192.59,138.81) --
	(194.72,138.61) --
	(196.85,138.85) --
	(198.98,138.54) --
	(201.11,138.96) --
	(203.24,138.32);

\path[draw=drawColor,line width= 1.1pt,dash pattern=on 4pt off 4pt ,line join=round] ( 45.65,138.66) --
	( 47.78,140.18) --
	( 49.91,145.55) --
	( 52.04,150.32) --
	( 54.17,152.97) --
	( 56.30,154.21) --
	( 58.43,154.89) --
	( 60.56,155.22) --
	( 62.69,155.41) --
	( 64.82,155.32) --
	( 66.95,154.98) --
	( 69.08,153.97) --
	( 71.21,151.93) --
	( 73.34,148.13) --
	( 75.47,141.95) --
	( 77.60,132.42) --
	( 79.73,119.13) --
	( 81.86,102.71) --
	( 83.99, 86.18) --
	( 86.12, 74.16) --
	( 88.25, 71.12) --
	( 90.37, 77.78) --
	( 92.50, 91.32) --
	( 94.63,107.17) --
	( 96.76,121.83) --
	( 98.89,133.33) --
	(101.02,141.58) --
	(103.15,146.82) --
	(105.28,149.94) --
	(107.41,151.35) --
	(109.54,151.93) --
	(111.67,151.78) --
	(113.80,151.46) --
	(115.93,150.64) --
	(118.06,149.98) --
	(120.19,148.83) --
	(122.32,150.99) --
	(124.45,152.49) --
	(126.58,149.55) --
	(128.71,147.05) --
	(130.84,145.35) --
	(132.97,144.52) --
	(135.10,143.81) --
	(137.22,143.36) --
	(139.35,142.86) --
	(141.48,142.51) --
	(143.61,142.11) --
	(145.74,141.85) --
	(147.87,141.52) --
	(150.00,141.35) --
	(152.13,141.06) --
	(154.26,141.00) --
	(156.39,140.56) --
	(158.52,142.14) --
	(160.65,150.32) --
	(162.78,149.72) --
	(164.91,144.28) --
	(167.04,140.99) --
	(169.17,139.43) --
	(171.30,138.97) --
	(173.43,138.76) --
	(175.56,138.76) --
	(177.69,138.70) --
	(179.82,138.73) --
	(181.94,138.70) --
	(184.07,138.73) --
	(186.20,138.71) --
	(188.33,138.73) --
	(190.46,138.71) --
	(192.59,138.72) --
	(194.72,138.71) --
	(196.85,138.72) --
	(198.98,138.72) --
	(201.11,138.71) --
	(203.24,138.74);

\path[draw=drawColor,line width= 1.1pt,dash pattern=on 1pt off 3pt ,line join=round] ( 45.65,138.72) --
	( 47.78,139.01) --
	( 49.91,140.15) --
	( 52.04,141.13) --
	( 54.17,141.68) --
	( 56.30,141.91) --
	( 58.43,142.07) --
	( 60.56,142.16) --
	( 62.69,142.32) --
	( 64.82,142.45) --
	( 66.95,142.67) --
	( 69.08,142.88) --
	( 71.21,143.19) --
	( 73.34,143.50) --
	( 75.47,143.89) --
	( 77.60,144.29) --
	( 79.73,144.77) --
	( 81.86,145.24) --
	( 83.99,145.78) --
	( 86.12,146.33) --
	( 88.25,147.00) --
	( 90.37,147.72) --
	( 92.50,148.58) --
	( 94.63,149.42) --
	( 96.76,150.27) --
	( 98.89,150.92) --
	(101.02,151.43) --
	(103.15,151.61) --
	(105.28,151.42) --
	(107.41,150.30) --
	(109.54,147.87) --
	(111.67,143.30) --
	(113.80,136.00) --
	(115.93,124.91) --
	(118.06,110.29) --
	(120.19, 93.34) --
	(122.32, 80.08) --
	(124.45, 73.16) --
	(126.58, 73.45) --
	(128.71, 82.74) --
	(130.84, 97.31) --
	(132.97,112.90) --
	(135.10,126.23) --
	(137.22,136.36) --
	(139.35,143.21) --
	(141.48,147.53) --
	(143.61,149.81) --
	(145.74,150.91) --
	(147.87,151.10) --
	(150.00,151.01) --
	(152.13,150.33) --
	(154.26,149.93) --
	(156.39,148.29) --
	(158.52,152.87) --
	(160.65,178.23) --
	(162.78,175.08) --
	(164.91,157.02) --
	(167.04,146.14) --
	(169.17,141.07) --
	(171.30,139.52) --
	(173.43,138.87) --
	(175.56,138.82) --
	(177.69,138.69) --
	(179.82,138.75) --
	(181.94,138.69) --
	(184.07,138.74) --
	(186.20,138.70) --
	(188.33,138.74) --
	(190.46,138.70) --
	(192.59,138.73) --
	(194.72,138.70) --
	(196.85,138.73) --
	(198.98,138.70) --
	(201.11,138.73) --
	(203.24,138.71);

\path[draw=drawColor,line width= 1.1pt,dash pattern=on 1pt off 3pt on 4pt off 3pt ,line join=round] ( 45.65,138.74) --
	( 47.78,138.79) --
	( 49.91,139.19) --
	( 52.04,139.46) --
	( 54.17,139.67) --
	( 56.30,139.69) --
	( 58.43,139.77) --
	( 60.56,139.75) --
	( 62.69,139.82) --
	( 64.82,139.81) --
	( 66.95,139.89) --
	( 69.08,139.88) --
	( 71.21,139.97) --
	( 73.34,139.98) --
	( 75.47,140.08) --
	( 77.60,140.10) --
	( 79.73,140.22) --
	( 81.86,140.25) --
	( 83.99,140.38) --
	( 86.12,140.43) --
	( 88.25,140.59) --
	( 90.37,140.66) --
	( 92.50,140.84) --
	( 94.63,140.94) --
	( 96.76,141.15) --
	( 98.89,141.28) --
	(101.02,141.54) --
	(103.15,141.71) --
	(105.28,142.04) --
	(107.41,142.29) --
	(109.54,142.71) --
	(111.67,143.03) --
	(113.80,143.55) --
	(115.93,143.93) --
	(118.06,144.55) --
	(120.19,144.94) --
	(122.32,146.70) --
	(124.45,148.16) --
	(126.58,148.13) --
	(128.71,148.20) --
	(130.84,148.73) --
	(132.97,149.44) --
	(135.10,150.22) --
	(137.22,150.77) --
	(139.35,151.23) --
	(141.48,151.22) --
	(143.61,150.77) --
	(145.74,149.05) --
	(147.87,145.90) --
	(150.00,140.07) --
	(152.13,131.38) --
	(154.26,118.39) --
	(156.39,102.75) --
	(158.52, 86.15) --
	(160.65, 88.20) --
	(162.78,109.81) --
	(164.91,126.25) --
	(167.04,134.44) --
	(169.17,137.26) --
	(171.30,138.39) --
	(173.43,138.55) --
	(175.56,138.75) --
	(177.69,138.66) --
	(179.82,138.76) --
	(181.94,138.67) --
	(184.07,138.76) --
	(186.20,138.68) --
	(188.33,138.75) --
	(190.46,138.68) --
	(192.59,138.75) --
	(194.72,138.69) --
	(196.85,138.75) --
	(198.98,138.69) --
	(201.11,138.74) --
	(203.24,138.69);
\definecolor{drawColor}{RGB}{255,255,255}

\path[draw=drawColor,line width= 0.5pt,line join=round,line cap=round] ( 37.78, 32.50) rectangle (211.12,185.17);
\end{scope}
\begin{scope}
\path[clip] (  0.00,  0.00) rectangle (216.81,195.13);
\definecolor{drawColor}{gray}{0.45}

\node[text=drawColor,anchor=base east,inner sep=0pt, outer sep=0pt, scale=  0.90] at ( 33.73, 43.59) {-15};

\node[text=drawColor,anchor=base east,inner sep=0pt, outer sep=0pt, scale=  0.90] at ( 33.73, 74.26) {-10};

\node[text=drawColor,anchor=base east,inner sep=0pt, outer sep=0pt, scale=  0.90] at ( 33.73,104.94) {-5};

\node[text=drawColor,anchor=base east,inner sep=0pt, outer sep=0pt, scale=  0.90] at ( 33.73,135.62) {0};

\node[text=drawColor,anchor=base east,inner sep=0pt, outer sep=0pt, scale=  0.90] at ( 33.73,166.30) {5};
\end{scope}
\begin{scope}
\path[clip] (  0.00,  0.00) rectangle (216.81,195.13);
\definecolor{drawColor}{gray}{0.45}

\node[text=drawColor,anchor=base,inner sep=0pt, outer sep=0pt, scale=  0.90] at ( 45.65, 22.25) {0};

\node[text=drawColor,anchor=base,inner sep=0pt, outer sep=0pt, scale=  0.90] at ( 89.27, 22.25) {10};

\node[text=drawColor,anchor=base,inner sep=0pt, outer sep=0pt, scale=  0.90] at (132.88, 22.25) {20};

\node[text=drawColor,anchor=base,inner sep=0pt, outer sep=0pt, scale=  0.90] at (176.49, 22.25) {30};
\end{scope}
\begin{scope}
\path[clip] (  0.00,  0.00) rectangle (216.81,195.13);
\definecolor{drawColor}{gray}{0.32}

\node[text=drawColor,rotate= 90.00,anchor=base,inner sep=0pt, outer sep=0pt, scale=  1.10] at ( 17.05,108.83) {SFAP [k $\cdot$ mV]};
\end{scope}
\end{tikzpicture}

    }
    \vspace*{-10.5mm}
    %\caption{}
    \label{fig:sfaps-a}
  \end{subfigure}
  \begin{subfigure}[t]{\columnwidth}
    \centering
    {\footnotesize
      % Created by tikzDevice version 0.12 on 2019-03-20 22:22:56
% !TEX encoding = UTF-8 Unicode
\begin{tikzpicture}[x=1pt,y=1pt]
\definecolor{fillColor}{RGB}{255,255,255}
\path[use as bounding box,fill=fillColor,fill opacity=0.00] (0,0) rectangle (216.81,195.13);
\begin{scope}
\path[clip] (  0.00,  0.00) rectangle (216.81,195.13);
\definecolor{drawColor}{RGB}{255,255,255}
\definecolor{fillColor}{RGB}{255,255,255}

\path[draw=drawColor,line width= 0.5pt,line join=round,line cap=round,fill=fillColor] (  0.00,  0.00) rectangle (216.81,195.13);
\end{scope}
\begin{scope}
\path[clip] ( 33.28, 32.50) rectangle (211.12,185.17);
\definecolor{drawColor}{RGB}{255,255,255}
\definecolor{fillColor}{RGB}{255,255,255}

\path[draw=drawColor,line width= 0.5pt,line join=round,line cap=round,fill=fillColor] ( 33.28, 32.50) rectangle (211.12,185.17);
\definecolor{drawColor}{gray}{0.85}

\path[draw=drawColor,line width= 0.3pt,line join=round] ( 33.28, 53.73) --
	(211.12, 53.73);

\path[draw=drawColor,line width= 0.3pt,line join=round] ( 33.28, 94.11) --
	(211.12, 94.11);

\path[draw=drawColor,line width= 0.3pt,line join=round] ( 33.28,134.50) --
	(211.12,134.50);

\path[draw=drawColor,line width= 0.3pt,line join=round] ( 33.28,174.88) --
	(211.12,174.88);

\path[draw=drawColor,line width= 0.3pt,line join=round] ( 41.36, 32.50) --
	( 41.36,185.17);

\path[draw=drawColor,line width= 0.3pt,line join=round] ( 86.11, 32.50) --
	( 86.11,185.17);

\path[draw=drawColor,line width= 0.3pt,line join=round] (130.85, 32.50) --
	(130.85,185.17);

\path[draw=drawColor,line width= 0.3pt,line join=round] (175.59, 32.50) --
	(175.59,185.17);
\definecolor{drawColor}{RGB}{0,0,0}

\path[draw=drawColor,line width= 1.1pt,line join=round] ( 41.36, 94.51) --
	( 43.55, 88.08) --
	( 45.73, 67.13) --
	( 47.91, 48.74) --
	( 50.10, 40.94) --
	( 52.28, 39.44) --
	( 54.47, 41.73) --
	( 56.65, 45.43) --
	( 58.84, 50.45) --
	( 61.02, 55.44) --
	( 63.21, 60.95) --
	( 65.39, 66.00) --
	( 67.58, 71.42) --
	( 69.76, 76.20) --
	( 71.95, 81.19) --
	( 74.13, 85.34) --
	( 76.32, 89.57) --
	( 78.50, 92.81) --
	( 80.69, 96.12) --
	( 82.87, 98.38) --
	( 85.06,100.80) --
	( 87.24,102.18) --
	( 89.43,103.87) --
	( 91.61,104.53) --
	( 93.80,105.70) --
	( 95.98,105.81) --
	( 98.17,106.66) --
	(100.35,106.37) --
	(102.53,107.07) --
	(104.72,106.44) --
	(106.90,107.13) --
	(109.09,106.11) --
	(111.27,107.01) --
	(113.46,105.38) --
	(115.64,106.96) --
	(117.83,104.19) --
	(120.01,135.32) --
	(122.20,158.90) --
	(124.38,135.44) --
	(126.57,115.94) --
	(128.75,104.09) --
	(130.94,100.88) --
	(133.12, 98.54) --
	(135.31, 98.68) --
	(137.49, 97.76) --
	(139.68, 98.13) --
	(141.86, 97.44) --
	(144.05, 97.79) --
	(146.23, 97.18) --
	(148.42, 97.52) --
	(150.60, 96.91) --
	(152.79, 97.39) --
	(154.97, 96.32) --
	(157.15,101.08) --
	(159.34,123.24) --
	(161.52,122.68) --
	(163.71,108.47) --
	(165.89,100.17) --
	(168.08, 95.86) --
	(170.26, 94.89) --
	(172.45, 94.11) --
	(174.63, 94.32) --
	(176.82, 93.97) --
	(179.00, 94.26) --
	(181.19, 93.97) --
	(183.37, 94.25) --
	(185.56, 93.97) --
	(187.74, 94.25) --
	(189.93, 93.97) --
	(192.11, 94.25) --
	(194.30, 93.96) --
	(196.48, 94.27) --
	(198.67, 93.94) --
	(200.85, 94.31) --
	(203.04, 93.86);

\path[draw=drawColor,line width= 1.1pt,dash pattern=on 4pt off 4pt ,line join=round] ( 41.36, 94.15) --
	( 43.55, 94.68) --
	( 45.73, 97.07) --
	( 47.91, 98.74) --
	( 50.10, 99.38) --
	( 52.28, 98.75) --
	( 54.47, 97.75) --
	( 56.65, 96.03) --
	( 58.84, 94.25) --
	( 61.02, 91.91) --
	( 63.21, 89.66) --
	( 65.39, 86.97) --
	( 67.58, 84.48) --
	( 69.76, 81.61) --
	( 71.95, 79.01) --
	( 74.13, 76.22) --
	( 76.32, 74.09) --
	( 78.50, 71.97) --
	( 80.69, 70.63) --
	( 82.87, 69.45) --
	( 85.06, 69.20) --
	( 87.24, 69.16) --
	( 89.43, 70.11) --
	( 91.61, 71.08) --
	( 93.80, 73.01) --
	( 95.98, 74.85) --
	( 98.17, 77.50) --
	(100.35, 79.65) --
	(102.53, 82.48) --
	(104.72, 84.57) --
	(106.90, 87.43) --
	(109.09, 89.25) --
	(111.27, 92.05) --
	(113.46, 93.25) --
	(115.64, 96.09) --
	(117.83, 96.25) --
	(120.01,115.11) --
	(122.20,129.63) --
	(124.38,117.81) --
	(126.57,107.91) --
	(128.75,102.08) --
	(130.94,100.86) --
	(133.12,100.02) --
	(135.31,100.43) --
	(137.49,100.18) --
	(139.68,100.57) --
	(141.86,100.30) --
	(144.05,100.59) --
	(146.23,100.24) --
	(148.42,100.49) --
	(150.60, 99.97) --
	(152.79,100.46) --
	(154.97, 98.99) --
	(157.15,107.15) --
	(159.34,147.28) --
	(161.52,145.62) --
	(163.71,120.24) --
	(165.89,104.87) --
	(168.08, 97.44) --
	(170.26, 95.35) --
	(172.45, 94.27) --
	(174.63, 94.32) --
	(176.82, 94.02) --
	(179.00, 94.22) --
	(181.19, 94.02) --
	(183.37, 94.20) --
	(185.56, 94.03) --
	(187.74, 94.19) --
	(189.93, 94.03) --
	(192.11, 94.19) --
	(194.30, 94.04) --
	(196.48, 94.18) --
	(198.67, 94.04) --
	(200.85, 94.17) --
	(203.04, 94.06);

\path[draw=drawColor,line width= 1.1pt,dash pattern=on 1pt off 3pt ,line join=round] ( 41.36, 94.12) --
	( 43.55, 95.07) --
	( 45.73, 98.83) --
	( 47.91,101.93) --
	( 50.10,103.67) --
	( 52.28,104.17) --
	( 54.47,104.45) --
	( 56.65,104.34) --
	( 58.84,104.40) --
	( 61.02,104.20) --
	( 63.21,104.21) --
	( 65.39,103.96) --
	( 67.58,103.91) --
	( 69.76,103.59) --
	( 71.95,103.46) --
	( 74.13,103.02) --
	( 76.32,102.73) --
	( 78.50,102.09) --
	( 80.69,101.57) --
	( 82.87,100.63) --
	( 85.06, 99.77) --
	( 87.24, 98.42) --
	( 89.43, 97.15) --
	( 91.61, 95.33) --
	( 93.80, 93.59) --
	( 95.98, 91.29) --
	( 98.17, 89.15) --
	(100.35, 86.45) --
	(102.53, 84.07) --
	(104.72, 81.15) --
	(106.90, 78.71) --
	(109.09, 75.63) --
	(111.27, 73.33) --
	(113.46, 70.40) --
	(115.64, 69.01) --
	(117.83, 66.49) --
	(120.01, 76.05) --
	(122.20, 83.72) --
	(124.38, 76.09) --
	(126.57, 70.29) --
	(128.75, 67.57) --
	(130.94, 68.17) --
	(133.12, 69.55) --
	(135.31, 72.09) --
	(137.49, 74.45) --
	(139.68, 77.29) --
	(141.86, 79.73) --
	(144.05, 82.54) --
	(146.23, 84.81) --
	(148.42, 87.52) --
	(150.60, 89.27) --
	(152.79, 92.12) --
	(154.97, 91.90) --
	(157.15,107.38) --
	(159.34,178.17) --
	(161.52,178.23) --
	(163.71,137.23) --
	(165.89,111.86) --
	(168.08, 99.69) --
	(170.26, 96.10) --
	(172.45, 94.44) --
	(174.63, 94.39) --
	(176.82, 94.02) --
	(179.00, 94.23) --
	(181.19, 94.01) --
	(183.37, 94.20) --
	(185.56, 94.03) --
	(187.74, 94.19) --
	(189.93, 94.04) --
	(192.11, 94.18) --
	(194.30, 94.05) --
	(196.48, 94.17) --
	(198.67, 94.06) --
	(200.85, 94.16) --
	(203.04, 94.08);

\path[draw=drawColor,line width= 1.1pt,dash pattern=on 1pt off 3pt on 4pt off 3pt ,line join=round] ( 41.36, 94.12) --
	( 43.55, 94.57) --
	( 45.73, 96.39) --
	( 47.91, 97.88) --
	( 50.10, 98.74) --
	( 52.28, 99.00) --
	( 54.47, 99.19) --
	( 56.65, 99.18) --
	( 58.84, 99.28) --
	( 61.02, 99.26) --
	( 63.21, 99.38) --
	( 65.39, 99.38) --
	( 67.58, 99.52) --
	( 69.76, 99.55) --
	( 71.95, 99.73) --
	( 74.13, 99.79) --
	( 76.32,100.00) --
	( 78.50,100.07) --
	( 80.69,100.31) --
	( 82.87,100.39) --
	( 85.06,100.64) --
	( 87.24,100.71) --
	( 89.43,100.95) --
	( 91.61,100.99) --
	( 93.80,101.21) --
	( 95.98,101.19) --
	( 98.17,101.39) --
	(100.35,101.30) --
	(102.53,101.46) --
	(104.72,101.27) --
	(106.90,101.39) --
	(109.09,101.05) --
	(111.27,101.12) --
	(113.46,100.52) --
	(115.64,100.60) --
	(117.83, 99.54) --
	(120.01,106.07) --
	(122.20,110.80) --
	(124.38,104.77) --
	(126.57, 99.27) --
	(128.75, 95.27) --
	(130.94, 92.90) --
	(133.12, 90.53) --
	(135.31, 88.44) --
	(137.49, 85.99) --
	(139.68, 83.60) --
	(141.86, 80.96) --
	(144.05, 78.39) --
	(146.23, 75.68) --
	(148.42, 73.03) --
	(150.60, 70.48) --
	(152.79, 68.25) --
	(154.97, 66.49) --
	(157.15, 65.16) --
	(159.34, 70.14) --
	(161.52, 81.14) --
	(163.71, 88.62) --
	(165.89, 92.28) --
	(168.08, 93.46) --
	(170.26, 93.99) --
	(172.45, 94.02) --
	(174.63, 94.14) --
	(176.82, 94.07) --
	(179.00, 94.15) --
	(181.19, 94.07) --
	(183.37, 94.15) --
	(185.56, 94.08) --
	(187.74, 94.14) --
	(189.93, 94.08) --
	(192.11, 94.14) --
	(194.30, 94.08) --
	(196.48, 94.14) --
	(198.67, 94.08) --
	(200.85, 94.14) --
	(203.04, 94.09);
\definecolor{drawColor}{RGB}{255,255,255}

\path[draw=drawColor,line width= 0.5pt,line join=round,line cap=round] ( 33.28, 32.50) rectangle (211.12,185.17);
\end{scope}
\begin{scope}
\path[clip] (  0.00,  0.00) rectangle (216.81,195.13);
\definecolor{drawColor}{gray}{0.45}

\node[text=drawColor,anchor=base east,inner sep=0pt, outer sep=0pt, scale=  0.90] at ( 29.23, 50.63) {-1};

\node[text=drawColor,anchor=base east,inner sep=0pt, outer sep=0pt, scale=  0.90] at ( 29.23, 91.01) {0};

\node[text=drawColor,anchor=base east,inner sep=0pt, outer sep=0pt, scale=  0.90] at ( 29.23,131.40) {1};

\node[text=drawColor,anchor=base east,inner sep=0pt, outer sep=0pt, scale=  0.90] at ( 29.23,171.78) {2};
\end{scope}
\begin{scope}
\path[clip] (  0.00,  0.00) rectangle (216.81,195.13);
\definecolor{drawColor}{gray}{0.45}

\node[text=drawColor,anchor=base,inner sep=0pt, outer sep=0pt, scale=  0.90] at ( 41.36, 22.25) {0};

\node[text=drawColor,anchor=base,inner sep=0pt, outer sep=0pt, scale=  0.90] at ( 86.11, 22.25) {10};

\node[text=drawColor,anchor=base,inner sep=0pt, outer sep=0pt, scale=  0.90] at (130.85, 22.25) {20};

\node[text=drawColor,anchor=base,inner sep=0pt, outer sep=0pt, scale=  0.90] at (175.59, 22.25) {30};
\end{scope}
\begin{scope}
\path[clip] (  0.00,  0.00) rectangle (216.81,195.13);
\definecolor{drawColor}{gray}{0.32}

\node[text=drawColor,anchor=base,inner sep=0pt, outer sep=0pt, scale=  1.10] at (122.20,  8.54) {Time [ms]};
\end{scope}
\begin{scope}
\path[clip] (  0.00,  0.00) rectangle (216.81,195.13);
\definecolor{drawColor}{gray}{0.32}

\node[text=drawColor,rotate= 90.00,anchor=base,inner sep=0pt, outer sep=0pt, scale=  1.10] at ( 17.05,108.83) {SFAP [k $\cdot$ mV]};
\end{scope}
\end{tikzpicture}

    }
    \vspace*{-6mm}
    %\caption{}
    \label{fig:sfaps-b}
  \end{subfigure}
  \caption{Upper graph: Simulated surface potentials evoked by a single firing muscle fiber (Single Fiber Action Potentials, SFAPs) as detected by four surface electrodes positioned above the NMJ (solid), above one of the two fiber ends (dash-dotted), and at two positions in between (dashed, dotted). The simulation model is due to Farina and Merletti (2001)~\cite{farina01}. Lower graph: Same as upper graph, but for a deeper fiber.}
  \label{fig:sfaps}
  \vspace*{-5mm}
\end{figure}

The contraction of a muscle fiber is triggered by an initial depolarisation of the muscle fiber membrane at the neuromuscular junction (NMJ), which then propagates along the fiber in the form of an intracellular action potential (IAP) at a speed of $3$--$5\si{m/s}$.~\cite{merletti04}.
%initiated by the exertion of a neural impulse by the corresponding motor neuron in the spinal cord. Upon arrival at the neuromuscular junction (NMJ), this neural impulse provokes a depolarisation of the muscle fiber membrane which then propagates along the fiber in the form of an intracellular action potential (IAP) at a speed of $3$--$5\si{m/s}$, i.e., much slower than the sources in EEG or sound analysis~\cite{merletti04}.
During the IAP's propagation along the fiber, the fiber represents a distributed current source and sink that evokes a change in the skin surface potential.
The electric field induced by this current source can be considered quasi-static~\cite{plonsey_barr}, i.e., changes are propagated nearly instantaneously from the fiber to the skin. Figure~\ref{fig:sfaps} shows a simulation of the surface potentials resulting from a single activation of two muscle fibers at different depths, as measured by surface electrodes at different locations along the fibers.
Notably, the increased damping induced by the detection of a deeper fiber affects the end-of-fiber artifacts much less than the propagating signal components. This is due to the spatial low-pass behavior of the volume conductor and the end-of-fiber artifacts' low spatial frequency~\cite{merletti04}.\footnote{The additional end-of-fiber signal at around $\SI{18}{ms}$ is due to the two fiber halves being of different lengths - at that point in time, the wave front travelling along the opposite direction from the NMJ has reached the other fiber end.}
This discrepancy between the damping of the different signal components with increasing distance can not be represented by means of an instantaneous mixing model. A convolutive model, on the other hand, can precisely represent this behavior when we consider the input signal $s(t)$ to be an impulse train, i.e., a digital signal indicating whether the motor neuron fires an impulse at instant $t$. This mathematical formulation allows to represent the distributed current source by means of a point source, while all information on the spatial behavior is contained in the coefficients of the mixing filter.

In the most general setting, a $\nsensor$-channel EMG signal can be modelled as the sum of the contributions of all Motor Units\footnote{A MU denotes the ensemble of a motor neuron and all muscle fibers innervated by that neuron. Note that these fibers always fire synchronously.} (MUs) belonging to the muscles of interest.
In terms of the Motor Unit Action Potentials (MUAPs), i.e., the changes in surface potential that a single contraction of a particular MU evokes at a particular electrode, this model can be formulated in discrete time as
\begin{equation}
  \label{eq:1}
  \emgvec(\tdis) = \muapmat\, \sourcevec(\tdis),
\end{equation}
with the measurement vector
\begin{equation}
  \label{eq:3}
  \emgvec(\tdis) = [\emgchan_1(\tdis), \ldots, \emgchan_\nsensor(\tdis)]^T \in \R^{\nsensor \times 1}, 
\end{equation}
the mixing matrix  
\begin{equation}
  \label{eq:2}
  \muapmat = 
  \begin{pmatrix}
    \muapvec_{1,1}^T & \cdots & \muapvec_{1,\nsource}^T \\
    \vdots & \ddots & \vdots \\
    \muapvec_{\nsensor,1}^T & \cdots & \muapvec_{\nsensor,\nsource}^T \\
  \end{pmatrix}
  \in \R^{\nsensor \times \filtlen\nsource},
\end{equation}
where $\muapvec_{\ia,\ib} = [\muapscal_{\ia,\ib, 0} \cdots \muapscal_{\ia,\ib,(\filtlen - 1)}]^T \in \R^{\filtlen \times 1}$ contains the time course of the length-$\filtlen$ MUAP evoked at sensor $\ia$ by motor unit $\ib$, and with the source vector
\begin{equation}
  \label{eq:4}
  \sourcevec(\tdis) = [\sourcesubvec_1^T(\tdis), \sourcesubvec_2^T(\tdis), \ldots, \sourcesubvec_\nsource^T(\tdis)]^T \in \R^{\filtlen\nsource \times 1}, \text{where}
\end{equation}
\begin{equation*}
\sourcesubvec_\ib(\tdis) = [\source_\ib(\tdis), \source_\ib(\tdis-1), \ldots, \source_\ib(\tdis-\filtlen+1)]^T \in \R^{\filtlen \times 1}.
\end{equation*}

In all practical applications, $\nsensor \ll \nsource$ and hence the problem is strongly underdetermined.
One therefore usually wishes to approximate the system by a much smaller number of sources $\nsourceapp \ll \nsource$.
This is equivalent to assuming
\begin{equation}
  \emgchan_\ia(\tdis) = \sum\limits_{\ib=1}^{\nsource} \muapvec_{\ia,\ib}^T \sourcesubvec_\ib(\tdis)
   \approx \sum\limits_{\ib=1}^{\nsourceapp} \muapvecapp_{\ia,\ib}^T \sourcesubvecapp_\ib(\tdis) \myforall{\ia, \tdis}.
  \label{eq:source_app}
\end{equation}
Now, on the one hand, due to the large spatial extension of the respiratory muscles, different MUs belonging to the same muscle can produce MUAPs of strongly varying wave shape, depending on the relative position of the MU and the recording electrode.
On the other hand, the muscles are relatively well-separated, in contrast to, e.g., forearm muscles.
Hence, the MUAPs of MUs belonging to one respiratory muscle may vary in shape, but they will all be rather different from those of the other respiratory muscles. This argument provides an informal justification of approximation~\eqref{eq:source_app} when considering the $\muapvecapp_{\ia,\ib}$ representative of all MUs belonging to a particular muscle, and the $\sourcesubvecapp_\ib(\tdis)$ a superposition of the source signals of all MUs represented by $\muapvecapp_{\ia,\ib}$.

%Figure~\ref{fig:mimo_system} shows the structure of the resulting multiple-input multiple-output (MIMO) system.
%\begin{figure}[hbt] %[hbt]
%  \begin{center}
%    \includegraphics[height=4cm]{./figures/bss_model_modif_pq2_PR.eps}
%    \caption{General setup for blind MIMO signal processing. Here, $h_{i,j} = \muapvec_{j,i}$.}
%    \label{fig:mimo_system}
%  \end{center}
%  \vspace*{-4mm}
%\end{figure}
For source separation, i.e., for eliminating the crosstalk between the channels, a suitable MIMO demixing system needs to be designed, yielding the output signals
\begin{equation}
\label{bkfirdemixmodel}
  \demixedchan_\ib(\tdis) = \sum\limits_{\ia=1}^{\nsensor} \demixvec_{\ib,\ia}^T \emgchanvec_\ia(\tdis),
\end{equation}
%or in matrix form
%\begin{align}
%  \label{eq:6}
%  \demixedvec(\tdis) &= \demixmat \, \muapmat \, \sourcevec(\tdis) \nonumber \\
%  &=
%  \begin{pmatrix}
%    \demixvec_{1,1}^T & \cdots & \demixvec_{1, \nsensor}^T \\
%    \vdots & \ddots & \vdots \\
%    \demixvec_{\nsource, 1}^T & \cdots & \demixvec_{\nsource, \nsensor}^T \\
%  \end{pmatrix} \,
%  \muapmat \, \sourcevec(\tdis),
%\end{align}
where $\demixvec_{\ia,\ib} \in \R^{\filtlendem \times 1}$ are the coefficients of the demixing filters.
%\footnote{The filter length $\filtlendem$ may differ from the filter length $\filtlen$ of the mixing system.} 
The $\nsensor \cdot \nsourceapp \cdot \filtlendem$ filter coefficients $\demix_{\ia,\ib,\kappa}, \kappa = 0, \dots, \filtlendem - 1$ are to be blindly estimated by the BSS algorithm.
It can be shown that for an ideal separation the number $\nsensor$ of sensors must be at least equal to the assumed number $\nsourceapp$ of sources \cite{book07}. Hence, in this study we choose $\nsensor = \nsourceapp$.

\section{TRINICON-based Blind Source Separation for Convolutive Mixtures}
\label{sec:algorithm}
The most common method for BSS is the so-called {\em Independent Component Analysis (ICA)},
for which several different algorithms have been developed in recent years. However, the original ICA model is based
on instantaneous mixtures (i.e., $L=1$) \cite{hyvar_book}. To cover convolutive mixtures, i.e., to blindly estimate all filter coefficients $w_{p,q,\kappa}$, our examinations are based on 
a more general approach.
The employed concept called TRINICON (TRIple-N ICA for CONvolutive mixtures) is a framework for broadband adaptive MIMO signal processing algorithms \cite{book03,book07} and allows the simultaneous exploitation of all fundamental statistical signal properties, namely nongaussianity, nonwhiteness and nonstationarity. For convolutive BSS these signal properties
are used to minimize the {\em mutual information} known from information theory, i.e., the crosstalk between
all output signals of the demixing system.
It can be shown that in contrast to the instantaneous case $L=1$, the convolutive case $L>1$ requires the simultaneous
exploitation of at least two of the three signal properties.
%More specifically, nongaussianity and/or nonstationarity is required in addition to nonwhiteness.
In this study, we apply a specially adapted and efficient realization of the concept using fast Fourier transforms (FFTs).

%Faltungsmixturen wurden in der Signalverarbeitungsliteratur in den letzten Jahren traditionell vor allem im Hinblick
%auf Akustik-Anwendungen, z.B. f\"ur Interfaces zur freien Spracheingabe, untersucht. 
%Deshalb wird in dieser Studie die EMG-Anwendung...

In addition to the modeling by convolutive mixtures, another challenge in the present biomedical application
is the significant level-difference (typically three orders of magnitude) between the ECG component and the relevant EMG signals. We therefore apply a two-step approach which consists first of a so-called {\em sphering}, followed by the actual source separation as described above. The sphering stage, consisting of an eigenvalue decomposition of the spatial correlation matrix and a power normalization, turns out to be a crucial step in handling this level imbalance so that it leads to a decisive numerical improvement of the subsequent separation task.

Both the sphering stage as well as the separation stage can be derived rigorously from the TRINICON framework.
In the present study, they are both carried out in an offline fashion, i.e., calculations are performed with the complete dataset already available.
The efficient frequency-domain realization shown here is based on the use of a multivariate probability density function so that cross-correlations between
all frequency components are taken into account and the commonly known permutation ambiguity in traditional narrowband frequency-domain BSS algorithms is avoided.
It follows directly from the generic broadband frequency-domain coefficient update equation (10.72) in \cite{book03} by approximating the constraint matrices
$\mathbf{L}$ and $\mathbf{G}_{\cdot}^{\cdot}$. The coupling of the frequency components is ensured by using a spherical multivariate Laplacian density.
(See also \cite{book07}, p.~130, second column in fig.~4.10, for a classification of this realization.)

%The spatial prewhitening step (sphering) can be understood as an efficient implementation of the Hessian matrix in the more general Newton-type coefficient update
%which was shown, e.g., in \cite{buchner_asilomar10}. It corresponds to an inherent power normalization w.r.t. the sources, which is crucial in our application.

{\small
\begin{breakablealgorithm}
\caption{Convolutive Blind Source Separation based on {\em Independent Vector Analysis}}
\begin{algorithmic}[1]
%\Procedure{ComputeDemixingFilter}{}
\Algphase{Spatial prewhitening}:
\BState Principal component analysis %(eigenvalue decomposition of the spatial correlation matrix):
\begin{eqnarray}
\nonumber
\mathbf{E}\mathbf{D}\mathbf{E}^{\mathrm{T}} &=& \hat{\mathrm{E}}\left\{ \mathbf{x}(n) \mathbf{x}^{\mathrm{T}}(n) \right\},
\,\,\, \mathrm{where} \nonumber \\
& & \,\,\, \mathbf{x}(n)=\left[ x_1(n), \ldots, x_P(n) \right]^{\mathrm{T}} \nonumber 
\end{eqnarray}
\BState Sphering:
\begin{equation}
\nonumber
\left[ x_1(n), \ldots, x_P(n) \right]^{\mathrm{T}} \leftarrow
\mathbf{E}\mathbf{D}^{-1/2}\mathbf{E}^{\mathrm{T}} \mathbf{x}(n)
\end{equation}
\vspace*{-7mm}
\Algphase{Main algorithm}:
\BState Short-time Fourier transform by windowing:
%\For{all input channels $p=1, \ldots, P$}
\Statex 
\begin{equation}
\nonumber
X_p^{(\nu)}(m) \Leftarrow x_p(n), \quad p=1,\ldots,P
\end{equation}
%\EndFor
\BState Centering:
\Statex \begin{equation}
\nonumber
X_p^{(\nu)}(m) \leftarrow X_p^{(\nu)}(m) - \frac{1}{N} \sum_{m'=0}^{N-1} X_p^{(\nu)}(m')
\end{equation}
\For{$\ell=1,\ldots,\ell_{\mathrm{max}}$}:
\State  Circular convolution to obtain (preliminary) outputs:
\begin{eqnarray}
%\nonumber
\mathbf{Y}^{(\nu)}(m)&=& \mathbf{W}^{(\nu) \ell-1} \mathbf{X}^{(\nu)}(m),
\,\,\, \mathrm{with}\,\,\, \nonumber \\
 & & \mathbf{X}^{(\nu)}(m)=\left[ X_1^{(\nu)}(m), \ldots, X_P^{(\nu)}(m) \right]^{\mathrm{T}}, \nonumber \\
 & & \mathbf{Y}^{(\nu)}(m)=\left[ Y_1^{(\nu)}(m), \ldots, Y_P^{(\nu)}(m) \right]^{\mathrm{T}} \nonumber 
\end{eqnarray}
\nonumber
\State  Computation of broadband normalization factors:
\begin{equation}
\nonumber
b_p(m) = \sqrt{\frac{1}{M} \sum_{\nu=0}^{M-1} |Y_p^{(\nu)}(m)|^2 }
\end{equation}
\State Multivariate score function:
\begin{equation}
\nonumber
{\boldsymbol{\Phi}}^{(\nu)}(m) = \left[ Y_1^{(\nu)}(m) /b_1(m), \, \ldots \, , Y_P^{(\nu)}(m) /b_P(m) \right]^\mathrm{T}
\end{equation}
\State  Coefficient update:
\begin{eqnarray}
\quad\mathbf{W}^{(\nu)\ell} &=& \mathbf{W}^{(\nu)\ell-1} + \mu  \nonumber \\
&\times & \mbox{\hspace{-5mm}} \left[ \mathbf{I} - \frac{1}{N} \sum_{m=0}^{N-1} \boldsymbol{\Phi}^{(\nu)}(m) \left(\mathbf{Y}^{(\nu)}(m)\right)^\mathrm{H} \right] \mathbf{W}^{(\nu)\ell-1} \nonumber 
\end{eqnarray}
\State  Minimum distortion principle:
\begin{equation}
\nonumber
\mathbf{W}^{(\nu)\ell} \leftarrow \mathrm{diag} \left\{ \left( \mathbf{W}^{(\nu)\ell} \right)^{-1} \right\} \mathbf{W}^{(\nu)\ell}
\end{equation}
\EndFor
\BState  Transformation of MIMO filter coefficients to time domain:
\begin{equation}
\nonumber
\mathbf{W}(n) \Leftarrow \mathbf{W}^{(\nu)\ell_{\mathrm{max}}}, \,\, \nu=0,\ldots,M-1
\end{equation}
\hrulefill
\Statex ($n=$ time index, $m=$ (temporal) block index, $\nu=$ frequency index, $p=$ channel index,
$N=$ number of blocks, $M=2\filtlendem=$ number of frequency bands, $P=$ number of channels,
$\mu=$ step size, $\ell=$ iteration index)
\end{algorithmic}
\label{alg:demixing}
\end{breakablealgorithm}
}
\vspace*{1mm}

\section{Experimental Results}
\label{sec:results}
\begin{figure}[tb]
  \centering
  
  % Labels...
  \psfrag{diaph1}[cc][cc]{\raisebox{7mm}{\footnotesize diaph1 [mV]}}
  \psfrag{diaph2}[cc][cc]{\raisebox{7mm}{\footnotesize diaph2 [mV]}}      %{\scriptsize diaph2}
  \psfrag{sterno}[cc][cc]{\raisebox{7mm}{\footnotesize intercost [mV]}}
  \psfrag{intercost}[cc][cc]{\raisebox{7mm}{\footnotesize sterno [mV]}}
  \psfrag{time [seconds]}[tc][tc]{\raisebox{-4mm}{\footnotesize Time [s]}}
  
  % Tick marks...
  \psfrag{934}[tc][tc]{}
  \psfrag{936}[tc][tc]{\raisebox{-2mm}{\scriptsize 936}}
  \psfrag{938}[tc][tc]{\raisebox{-2mm}{\scriptsize 938}}
  \psfrag{940}[tc][tc]{\raisebox{-2mm}{\scriptsize 940}}
  \psfrag{942}[tc][tc]{\raisebox{-2mm}{\scriptsize 942}}
  \psfrag{944}[tc][tc]{\raisebox{-2mm}{\scriptsize 944}}
  \psfrag{946}[tc][tc]{\raisebox{-2mm}{\scriptsize 946}}
  \psfrag{948}[tc][tc]{\raisebox{-2mm}{\scriptsize 948}}
  \psfrag{950}[tc][tc]{}
  \psfrag{500}[cr][cr]{\scriptsize 500}
  \psfrag{200}[cr][cr]{\scriptsize 200}
  \psfrag{100}[cr][cr]{\scriptsize 100}
  \psfrag{50}[cr][cr]{\scriptsize 50}
  \psfrag{0}[cr][cr]{\scriptsize 0}
  \psfrag{-50}[cr][cr]{\scriptsize -50}
  \psfrag{-100}[cr][cr]{\scriptsize -100}
  \psfrag{-200}[cr][cr]{\scriptsize -200}
  \psfrag{-400}[cr][cr]{\scriptsize -400}
  \psfrag{-500}[cr][cr]{\scriptsize -500}
  \psfrag{-600}[cr][cr]{\scriptsize -600}
  \psfrag{-800}[cr][cr]{\scriptsize -800}
  \psfrag{-1000}[cr][cr]{\scriptsize -1000}
  \psfrag{-1500}[cr][cr]{\scriptsize -1500}
  
  \centerline{\epsfig{figure=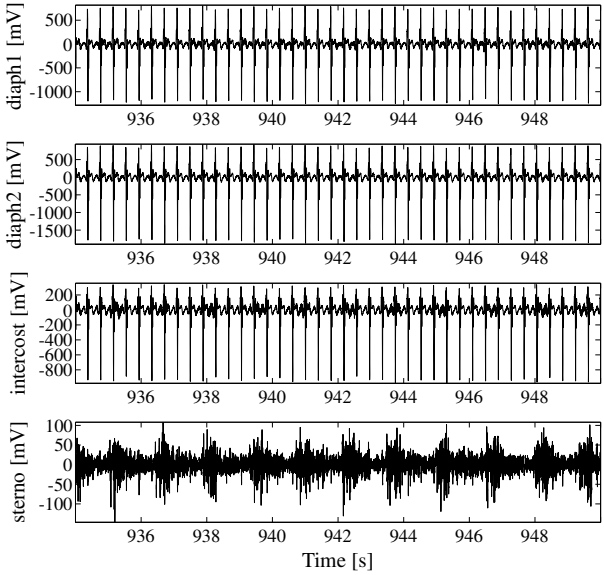,width=9cm}}
  \vspace*{-4mm}
  \caption{Measurement data consisting of four differentially recorded signals: {\em diaph1} (lower diaphragm), {\em diaph2} (upper diaphragm), {\em intercost} (intercostal muscles), and {\em sterno} (M. Scelenus / Sternocleido Mastoideus, on the neck). DC components were suppressed by highpass filtering. Sampling interval $T_\mathrm{a}=\SI{0.976}{ms}$.}
  \label{fig:sensordata}
  \vspace*{-4mm}
\end{figure}

\begin{figure}[tb]
  \centering
  \psfrag{output channel 1}[cc][cc]{\raisebox{4mm}{\scriptsize Channel 1 [au]}}
  \psfrag{output channel 2}[cc][cc]{\raisebox{4mm}{\scriptsize Channel 2 [au]}}      %{\scriptsize diaph2}
  \psfrag{output channel 3}[cc][cc]{\raisebox{4mm}{\scriptsize Channel 3 [au]}}
  \psfrag{output channel 4}[cc][cc]{\raisebox{4mm}{\scriptsize Channel 4 [au]}}
  \psfrag{time [seconds]}[tc][tc]{\raisebox{-4mm}{\footnotesize Time [s]}}
  
  % Tick marks...
  \psfrag{934}[tc][tc]{}
  \psfrag{936}[tc][tc]{\raisebox{-2mm}{\scriptsize 936}}
  \psfrag{938}[tc][tc]{\raisebox{-2mm}{\scriptsize 938}}
  \psfrag{940}[tc][tc]{\raisebox{-2mm}{\scriptsize 940}}
  \psfrag{942}[tc][tc]{\raisebox{-2mm}{\scriptsize 942}}
  \psfrag{944}[tc][tc]{\raisebox{-2mm}{\scriptsize 944}}
  \psfrag{946}[tc][tc]{\raisebox{-2mm}{\scriptsize 946}}
  \psfrag{948}[tc][tc]{\raisebox{-2mm}{\scriptsize 948}}
  \psfrag{950}[tc][tc]{}
  \psfrag{3}[cr][cr]{}
  \psfrag{2}[cr][cr]{\scriptsize 2}
  \psfrag{1}[cr][cr]{\scriptsize 1}
  \psfrag{0}[cr][cr]{\scriptsize 0}
  \psfrag{-1}[cr][cr]{\scriptsize -1}
  \psfrag{-2}[cr][cr]{\scriptsize -2}
  \psfrag{-3}[cr][cr]{}  
  \psfrag{-4}[cr][cr]{\scriptsize -4}
  
  \centerline{\epsfig{figure=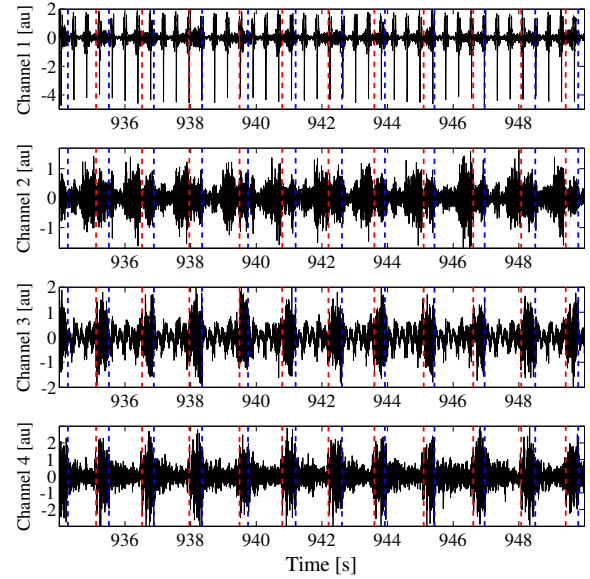,width=9cm}}
  \vspace*{-4mm}
  \caption{Output signals of the convolutive blind source separation. Dashed red lines indicate the beginning of inspiration; dashed blue lines the beginning of expiration. These have been derived manually from synchronously recorded pneumatic data. $au=$ arbitrary units.}
  \label{fig:bss_outputs}
  \vspace*{-4mm}
\end{figure}

The BSS algorithm described in the previous section has been applied to actual sEMG measurement data.\footnote{Data collected by Philippe Jolliet and his team during a clinical study entitled ``Comparative Effects on Diaphragmatic Electrical Activity and Respiratory Pattern of Various Levels of Assistance'', approved by the local ethics committee at Lausanne University Hospital (identifier: NCT01248845).} Figure~\ref{fig:sensordata} shows the four EMG channels that have been recorded - note the extremely dominant ECG component in the upper three channels.
Figure~\ref{fig:bss_outputs} shows the output signals generated by the BSS algorithm.
As desired, the ECG component (channel~1) seems to be reasonably well separated from the other signal sources.
Moreover, output channel~2 seems to capture expiratory muscle activity - supposedly from the abdominal muscles - while channels~3 and 4 show inspiratory muscle activity.
One might furthermore speculate that channels~3 and 4 most likely represent intercostal and diaphragmatic activity.
Note that there is no reference signal that could be used to validate the separated source signals (except for pneumatic signals, as shown), since the real level of activity of the different muscles is unknown.

The necessity of convolutive mixtures could also be confirmed experimentally on this data set.
For the results shown here, we empirically determined a necessary filter length of $L=64$.\footnote{The filter length was varied in powers of two in order to obtain an efficient BSS implementation using FFTs.}
In particular, it turned out that in the simple case of an instantaneous mixing model (i.e., $L=1$) it was not possible to obtain any significant separation.
Note that the filter length $L=64$ appears reasonable from a physiological point of view:
Each demixing filter effectively represents a propagation path of total length $\Delta s = \velocity L T_\mathrm{a} \approx \SI{25}{cm}$ (assuming $\velocity = \SI{4}{m/s}$, see above).
This is consistent with the dimensions of the muscle-sensor setup.% according to fig.~\ref{fig:elektrodenanordnung}.

\section{Conclusion}
\label{sec:conclusion}
In this article we have presented evidence that the use of convolutive BSS may significantly advance the preprocessing of surface EMG measurements of respiratory muscles.
Despite the limited validation data set, it can be concluded that the proposed algorithm has the potential of separating activity between inspiratory and expiratory muscles and removing ECG artifacts, both significantly better than previously proposed algorithms for instantaneous BSS.
While in the present study an offline algorithm was used to compute the coefficients of the demixing filter, an adaptive variant of the proposed algorithm is immanent: instead of only computing the coefficients of the demixing filter initially before performing the actual demixing, an online update of the filter coefficients using block or incremental updates can be performed, allowing for real-time applications. 
The performance of the proposed algorithm will be evaluated on simulated data as well as on a wider set of clinical data in a subsequent study.
%Another unexploited feature is the use of prior information. While the signals it self are not known a priori it statistical properties are. This information can be utilized in the computation of the de-mixing coefficients e.g. by considering an approach known as factor analysis.
%Finally, the efficiency of the algorithm may be improved by taking the expected high degree of sparsity in the demixing coefficients into account and introducing an additional $l_1$ regularization term into the calculation of the demixing coefficients.

\addtolength{\textheight}{-12cm}   % This command serves to balance the column lengths

\section*{Acknowledgments}
We thank Philippe Jolliet and his team at the Lausanne University Hospital for providing the clinical data used for the evaluation of the proposed algorithm.

\end{document}